\documentclass[preprint,showpacs,preprintnumbers,amsmath,amssymb]{revtex4}

\usepackage{graphicx}
\usepackage{dcolumn}
\usepackage{bm}

\begin{document}

\preprint{APS/123-QED}

\title{Numerical study of the antiferromagnetic Ising model on a hypersphere}

\author{
N.Olivi-Tran}
\affiliation{
SPCTS, UMR 6638, ENSCI, 47 avenue Albert Thomas, 87065 Limoges cedex,
France}

\author{R.V.Paredes}
\affiliation{
Grupo de Fisica Estadistica, Centro de Fisica, IVIC,  Apartado Postal 21827, Caracas 1020-A, Venezuela}

\date{\today}

\begin{abstract}
We built a model where all spins are in interaction  with each other via an antiferromagnetic
Ising Hamiltonian.
The geometry of such a model is a tetrahedron placed on a hypersphere
in spaces of dimensions enclosed between 1 and 9.
Due to confinement and to the fact that all spins interact
which each other, our spin system exhibit   frustration.
The temperatures of the observed antiferro-paramagnetic transitions
are equal for all space dimensions to one of two given values depending
on the parity of the space dimension.
Moreover, the order parameter $<m>$, i.e. the magnetization of the system,
has been also studied.

\end{abstract}

\maketitle
The Ising model has been widely studied to understand physical phenomena that occur
in ferro or antiferroelectric compounds, lattice gas, binary alloys etc. But, up to now, there are very few articles
on confinement effects \cite{binder,diehl,dosch,drze}  on the Ising model.
Understanding the statistical mechanics of classical systems in confined geometries
and in systems of small sizes is important for the future studies of nanocompounds.

This article deals of $N$ Ising spins on a $d=N-2$ dimensional hypersphere.
The geometry of the  spins corresponds to the following: each spin  is located on the apex of the tetrahedron
 corresponding to the space dimension.
 For the sake of confinement,
we assume that the space is closed: it corresponds to a $d$ dimensional hypersphere
 on which the tetrahedron is located. This tetrahedron is located
on the hypersphere so as each spin is at equal distance from each other.
Hence the curvature of the hypersphere is very large and is directly
related to the distance between spins.
The number of apex of the tetrahedron is also directly related to
the dimension of space: if $d$ is the dimension of  hyperspace,
the number of apex is equal to $d+2$.
So, like in a 1-dimensional circle or a 2-dimensional sphere,
each spin located at an apex of the tetrahedron is at equal distance
from all other spins.

 Hence the Hamiltonian of such an Ising model
writes:
\begin{equation}
H=-\sum _i \sum _{j \neq i} -J s_i s_j
\end{equation}
were $s_i$ is Ising spin number $i$ and $J<0$ is the antiferromagnetic coupling.
We took here the magnetisation $m=<|s_i|>$ of the assembly of spins as the order
 parameter
in a finite size analysis. The algorithm used here was the Wolff one.

The finite size analysis \cite{40} is a very efficient way to study
phase transitions by Monte-Carlo simulations. Indeed, the notion of phase transi
tion
has a sense only for the thermodynamical limit, while simulations can only be done
on finite size systems.
For the case of second order phase transitions, for infinite size systems with
periodic boundary conditions, the correlation length diverges at the critical
temperature $T_c$.
Here, we have a closed and finite space.

We shall define here the physical parameters necessary to the finite size analysis.
The specific heat per spin $c$ writes:
\begin{equation}
c(t)=\frac{<E^2>-<E>^2}{Nk_BT^2}
\end{equation}
where $E$ is the total energy of the assembly of spins, $T$ is the the absolute temperature, $k_B$ is Boltzmann constant and $t=T/T_c-1$ where $T_c$
is the critical temperature.
For the order parameter per spin $m$ we have:
\begin{equation}
m(t)=\frac{<|M|>}{N}
\end{equation}
where $M$ is the magnetization of the whole assembly of spins.
We took $J=1$ and $k_B=1$. The results are the following.

\begin{figure}[t]
\includegraphics[width=54mm]{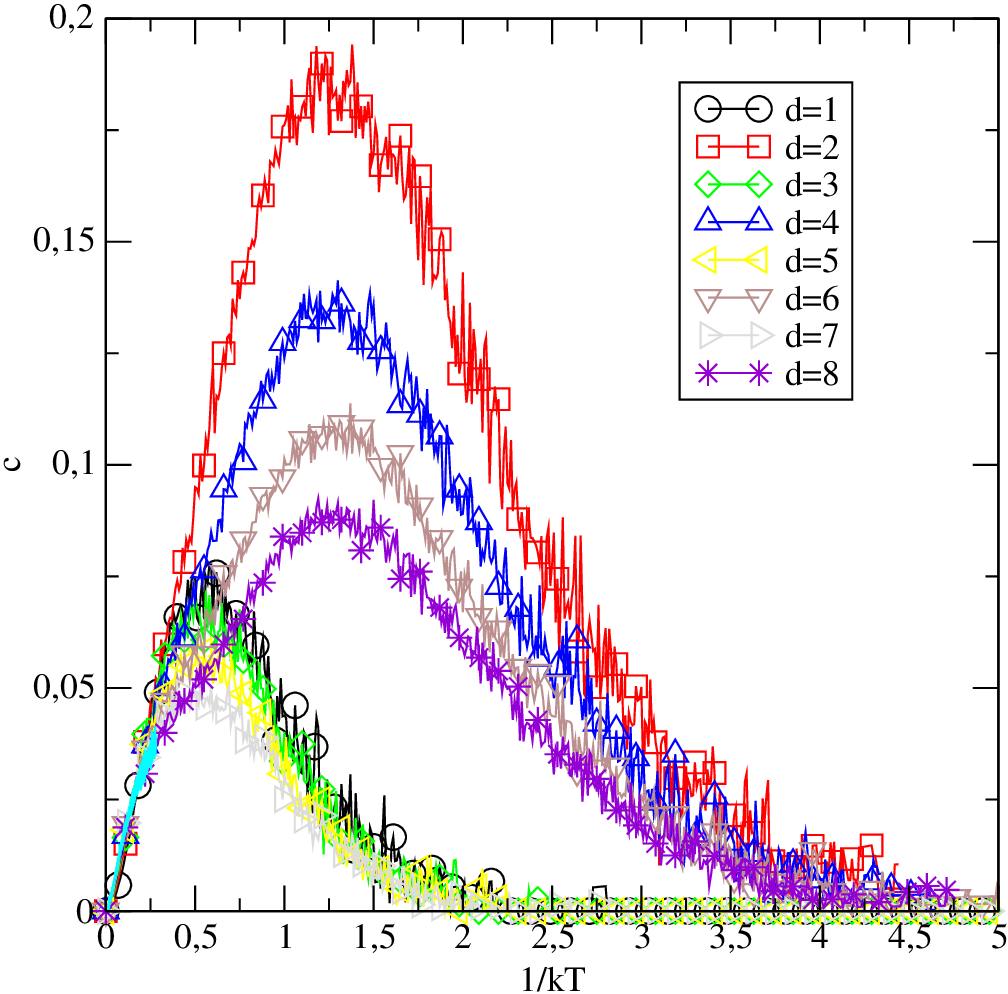}
\caption{Evolution of the specific heat $c$ as a function
of $\beta=1/k_B T$ for space dimensions : $d=1,2,3,4,5,6,7,8,9$}
\end{figure}

In fig.1, one can see the evolution of the specific heat $c$ as a function
of $\beta$.
For low space dimensions $d<10$ it appears a smooth maximum in this
graph which can be interpreted as a transition at a critical
temperature $\beta_c$.
 It is easy to see that if  the curved space has a dimension
which is odd the inverse of the critical temperature is equal to
$\beta_c=0.6$; if the hypersphere has an even space dimension
the critical temperature is equal to $\beta_c=1.25$.
This is valid for all space dimensions enclosed between $d=1$
and $d=9$.
But for high dimensions, the evolution of the peak of $c$ seems to flatten
as a function of $\beta$. Though, we have checked that transitions  always  occur  for
$N \rightarrow \infty$ and that the two values of $\beta_c$ remain.
Let us look now at fig.2 which is the evolution of the order parameter
$<m>$ as a function of the inverse temperature $\beta=1/k_BT$.
For $d=2$ and $d=4$, $<m>$ tends to zero when $\beta$ tends to infinity.
For the space dimension $d=1,3,5,6,7,8,9$ even for $\beta \rightarrow \infty$
the magnetization $<m>$ does not tend to zero.

\begin{figure}[h]
\includegraphics[width=54mm]{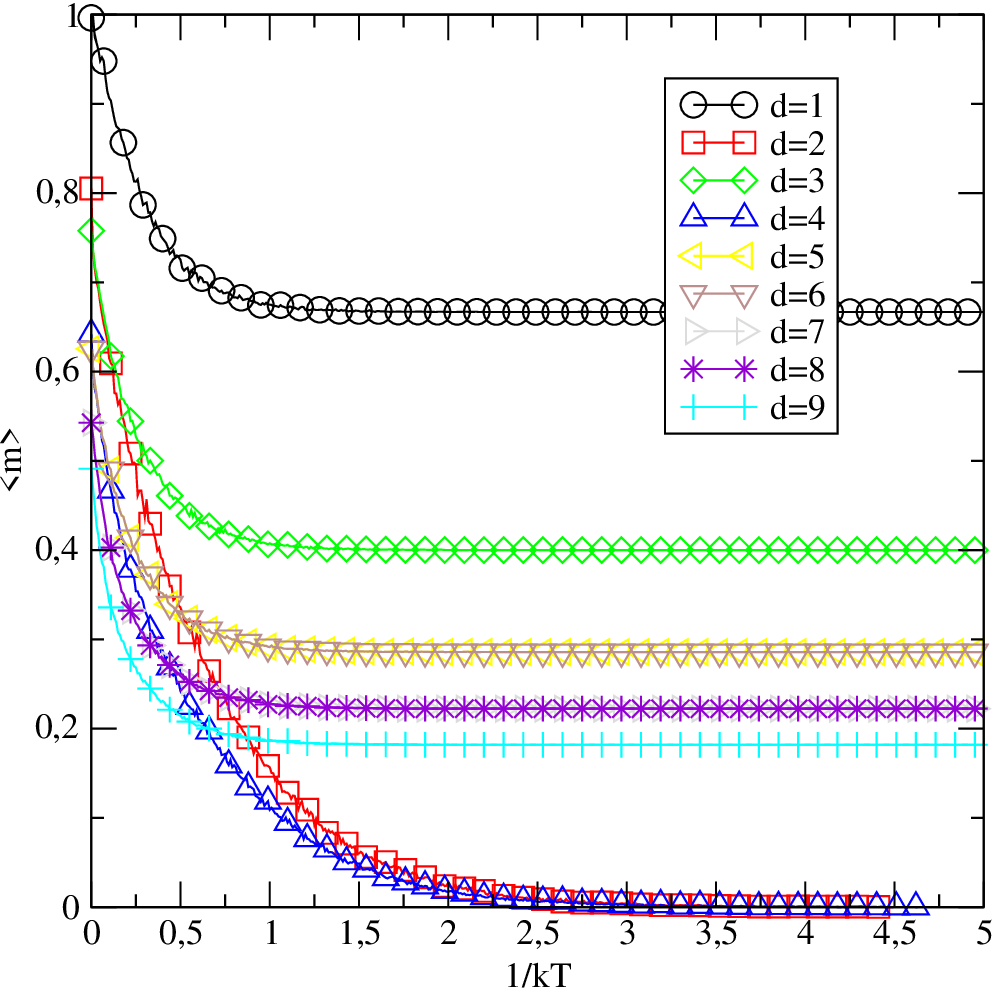}
\caption{Evolution of the order parameter $<m>$ as a function
of $\beta=1/k_B T$ for space dimensions : $d=1,2,3,4,5,6,7,8,9$}
\end{figure}

To resume fig.2 we plotted in fig. 3 the values of $<m>$ at $\beta=0$
(squares) and at $\beta=5$ (circles).
For $d=100$  we add the value of $<m>=0.156$ at $\beta=0$ and $<m>=0$ at $\beta=5$.

\begin{figure}[b]
\includegraphics[width=54mm]{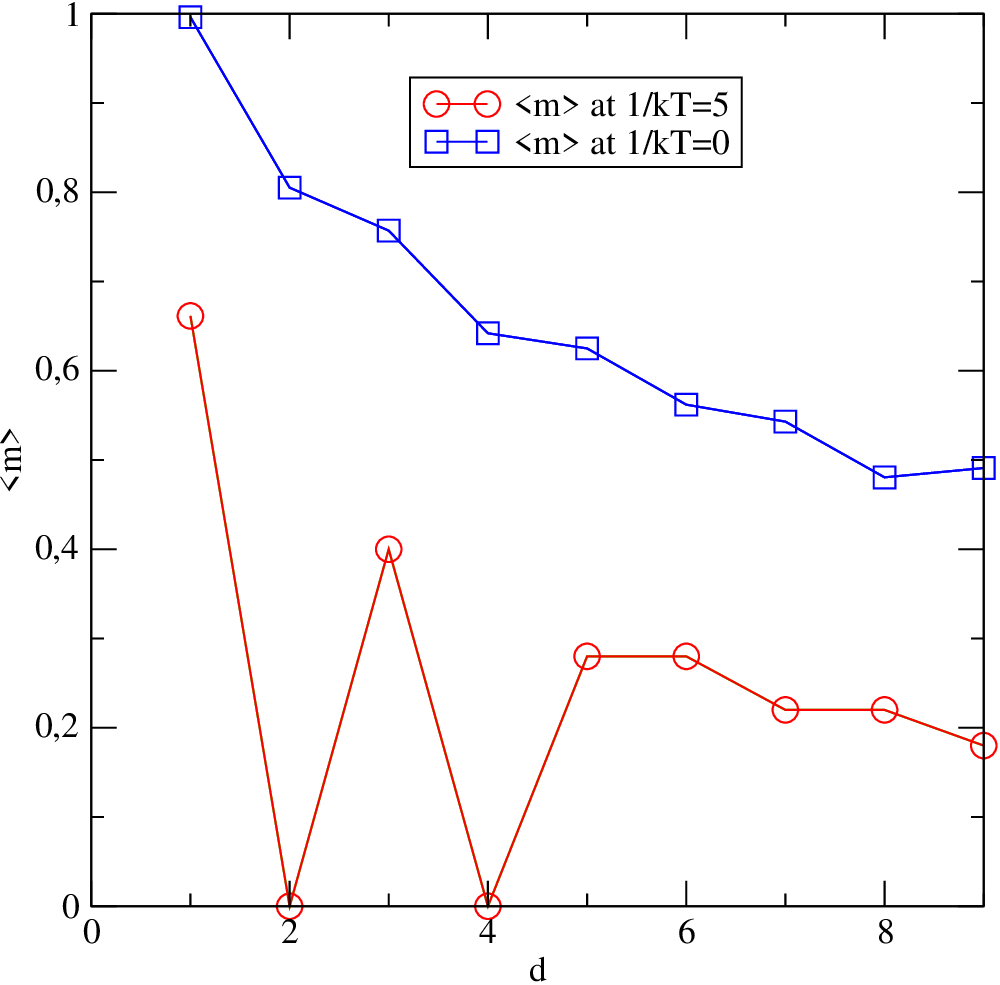}
\caption{Evolution of the $<m>$ at $1/k_BT=5$ (circles) and at $1/k_BT=0$
(squares) as a function of space dimension $d$}
\end{figure}

Geometrical frustration \cite{mosseri} is the explanation of the behavior of
both $<m>$ and $c$ as a function of $\beta$.
 $<m>$ at $1/k_BT=5$ is equal to zero for $d=2$ and for
$d=4$ because the number of antiferromagnetic spins is even in such spaces
and geometrical frustration is not too strong. Hence it is  possible
that the number of spins up and the number of spins down
is equal even if geometrical frustration remains because
of the tetrahedral geometry. For even space dimensions larger than 4,
geometrical frustration is too strong, and the number of spins up
and down is different leading to a non zero value of the magnetization
as the temperature goes to zero (i.e. $\beta \rightarrow \infty$).

If the space dimension is even, the number of spins is also even,
hence the transition from an antiferromagnetic state
to a paramagnetic one will occur at a lower temperature, i.e.
at a larger $\beta$ value because each spin has $d+1$ neighbours.
Hence, the number of interactions between spins is odd
and  $<E>^2$ always differs from $<E^2>$ of one value of spin interaction.
For odd space dimensions, the even number of interactions between spins renders it possible that $<E^2>$ is close to $<E>^2$, transitions occur at a larger temperature
(i.e. lower $\beta$).

The values of the critical temperature are the same
for all odd (resp. even) space dimensions: the tetrahedral geometry
is the same in all dimensions, no critical dimension has been observed
because of the curvature of space.
Moreover, we don not observe, as the number of spins increases, that
the critical value $\beta_c$ tends to zero, contrarily to large systems
in a threedimensional space.

This work has been done with the financial support
of a CNRS-FONACIT contract.

\end{document}